\documentclass[12pt,a4paper]{article}
\begin{document}
\large

\newpage
\begin{center}
{\bf MASS OF THE NEUTRINO AND ITS AXIAL - VECTOR
ELECTROMAGNETIC NATURE}
\end{center}
\vspace{1cm}
\begin{center}
{\bf Rasulkhozha S. Sharafiddinov}
\end{center}
\vspace{1cm}
\begin{center}
{\bf Institute of Nuclear Physics, Uzbekistan Academy of Sciences,
Tashkent, 702132 Ulugbek, Uzbekistan}
\end{center}
\vspace{1cm}

The neutrino possesses the anapole and electric dipole moments. Their
interaction with field of emission can also lead to the neutrino elastic
scattering by spinless nuclei. In this letter we present some implications
implied from the processes cross sections. One of them states that there
exists a hard connection between the neutrino magnetic and anapole moments.
The equation for the anapole and electric dipole form factors is also
obtained. They define the electronic neutrino axial - vector moments.
All findings are generalized to the case of a Majorana neutrino.

\newpage
According to the standard electroweak theory of elementary particles, the
neutrinos are strictly massless. At the same time a consistent theoretical
generalization of the $SU(2)_{L}$$\otimes$$U(1)$ model predicts the existence
of a massive Dirac neutrino. Herewith the neutrino interaction with virtual
photon is described by the vertex operator $\Gamma_{\mu}$ containing the
vector $\Gamma_{\mu}^{V}$ and axial - vector $\Gamma_{\mu}^{A}$ parts:
\begin{equation}
\Gamma_{\mu}(p,p')=\Gamma_{\mu}^{V}(p,p')+\Gamma_{\mu}^{A}(p,p'),
\label{1}
\end{equation}
\begin{equation}
\Gamma_{\mu}^{V}(p,p')=\overline{u}(p',s',)[\gamma_{\mu}F_{1\nu}(q^{2})-
i\sigma_{\mu\lambda}q_{\lambda}F_{2\nu}(q^{2})]u(p,s),
\label{2}
\end{equation}
\begin{equation}
\Gamma_{\mu}^{A}(p,p')=\overline{u}(p',s')\gamma_{5}[\gamma_{\mu}q^{2}
G_{1\nu}(q^{2})-i\sigma_{\mu\lambda}q_{\lambda}G_{2\nu}(q^{2})]u(p,s).
\label{3}
\end{equation}
Here $\sigma_{\mu\lambda}=[\gamma_{\mu},\gamma_{\lambda}]/2,$ $q=p-p',$
$p(s)$ and $p'(s')$ denote the four - momentum (helicity) of the neutrino
before and after emission. The functions $F_{1\nu}(q^{2}),$ $F_{2\nu}(q^{2}),$
$G_{1\nu}(q^{2})$ and $G_{2\nu}(q^{2})$ define at $q^{2}$=0 the static size of
the neutrino charge \cite{1}, magnetic \cite{2}, anapole \cite{3} and electric
dipole \cite{4} moments:
\begin{equation}
e_{\nu}=-F_{1\nu}(0), \, \, \, \, \mu_{\nu}=F_{2\nu}(0),
\label{4}
\end{equation}
\begin{equation}
{\it a_{\nu}}=-G_{1\nu}(0), \, \, \, \, d_{\nu}=G_{2\nu}(0),
\label{5}
\end{equation}
from which ${\it a_{\nu}}$ also can be measured experimentally \cite{5},
and for $e_{\nu},$ $\mu_{\nu}$ and $d_{\nu}$ was found the laboratory \cite{6}
and cosmological \cite{7} limits. However, the crucial value has for us a
question of whether there exists any dependence of form factors themselves.

The observation of such a regularity would help to elucidate the nature
of neutrinos of the different structure \cite{8}. All they therefore was
discussed in the present work studying the behavior of unpolarized and
longitudinal polarized electrons and their neutrinos at the elastic scattering
on a spinless nucleus arising because of the existence of fermions anapole and
electric dipole moments. Starting from the exactly formulas for the processes
cross sections, the equation between the anapole and electric dipole form
factors of light leptons have been established. A connection of the magnetic
moment and anapole is obtained. They state that if the neutrino corresponds
to the electron $(\nu=\nu_{e}),$ the functions $G_{1\nu}(0)$ and $G_{2\nu}(0)$
in the framework of the $(V-A)$ version of electroweak theory must have
the form
\begin{equation}
G_{1\nu}(0)=\frac{3eG_{F}}{8\pi^{2}\sqrt{2}}, \, \ \, \,
G_{2\nu}(0)=\frac{3eG_{F}m_{\nu}}{4\pi^{2}\sqrt{2}}.
\label{6}
\end{equation}
Here and further $e>0.$

Using (\ref{5}), (\ref{6}) and taking \cite{9}
$G_{F}=1.16637\cdot 10^{-5}\ {\rm GeV^{-2}},$
for axial - vector moments
of the neutrino with mass \cite{10} $m_{\nu}=10\ {\rm eV},$ we find
\begin{equation}
{\it a_{\nu}}=\frac{3G_{F}m_{e}}{4\pi^{2}\sqrt{2}}\ {\rm \mu_{B}}=
3.2\cdot 10^{-19}\ {\rm \mu_{B}}\left(\frac{1}{1\ {\rm eV}}\right),
\label{7}
\end{equation}
\begin{equation}
d_{\nu}=
6.268\cdot 10^{-25}\left(\frac{m_{\nu}}{1\ {\rm eV}}\right)\ {\rm e\cdot cm}=
6.27\cdot 10^{-24}\ {\rm e\cdot cm},
\label{8}
\end{equation}
where $\mu_{B}=e/2m_{e}$ is the electron Dirac magnetic moment.

It is seen that ${\it a_{\nu}}$ at the recent state of the theory do not
depend on the neutrino mass. At the same time the structural multipliers
$G_{F}$ and $m_{e}$ were measured with sufficiently exactness. Therefore,
to the estimate of (\ref{7}) one must apply simultaneously as to the
laboratory one. Insofar as the parameter $d_{\nu}$ is concerned, its value
becomes, according to the experimental data \cite{11}, equal to \cite{12}
$d_{\nu}< 0.44\cdot 10^{-20}\ {\rm e\cdot cm}.$ On the other hand the
cosmological reasoning \cite{7} give the justification that
$d_{\nu}< 2.5\cdot 10^{-22}\ {\rm e\cdot cm}.$ Such a bound closely
to the size of (\ref{8}), but the difference therein still exists.

Passing to the question about the Majorana neutrino \cite{13}, one can
as a starting recall \cite{14} that a truly neutral neutrino do not have
the vector interaction, and its axial - vector interaction is stronger
than the Dirac fermions.

Our analysis shows that a massive Majorana neutrino similarly to the
Dirac neutrino must possess not only one of the anapole \cite{15} or the
electric dipole \cite{16} moments, but each of them. We can present their
in the form
\begin{equation}
G_{1\nu_{M}}(0)=\frac{3eG_{F}}{4\pi^{2}\sqrt{2}}, \, \, \, \,
G_{2\nu_{M}}(0)=\frac{3eG_{F}m_{\nu_{M}}}{2\pi^{2}\sqrt{2}}.
\label{9}
\end{equation}

These functions together with form factors (\ref{6}) reflect just some
structural properties of all the masses of fermions.


\newpage
\begin{thebibliography}{99}
\bibitem{1} R.S. Sharafiddinov, Dokl. Akad. Nauk Ruz. Ser. Math.
Tehn. Estest. {\bf 7}, 25 (1998).
\bibitem{2} K. Fujikawa and R.E. Shrock, Phys. Rev. Lett. {\bf 45},
963 (1980).
\bibitem{3} Ya.B. Zel'dovich, Zh. Eksp. Teor. Fiz. {\bf 33}, 1531 (1957);
Ya.B. Zel'dovich and A.M. Perelomov, Zh. Eksp.
Teor. Fiz. {\bf 39}, 1115 (1960).
\bibitem{4} R.B. Begzhanov and R.S. Sharafiddinov, Mod. Phys.
Lett. {\bf A 15}, 557 (2000); Izv. Russ. Acad. Nauk Ser.
Fiz. {\bf 64}, 2221 (2000).
\bibitem{5} M.J. Musolf and B.R. Holstein, Phys. Rev.
{\bf D 43}, 1956 (1991).
\bibitem{6} S. Davidson, B. Campbell and K.D. Bailey, Phys.
Rev. {\bf D 43}, 2314 (1991).
\bibitem{7} J.A. Morgan and D.B. Farrant, Phys. Lett.
{\bf B 128}, 431 (1983).
\bibitem{8} R.B. Begzhanov and R.S. Sharafiddinov, in {\it Proc. Int.
Conf. on Nuclear Physics}, Moscow, June 16-19, 1998 (St-Petersburg,
1998), Abstracts, p.353
\bibitem{9} G. Bardin et al., Phys. Lett. {\bf B 137}, 135 (1984).
\bibitem{10} R.G.H. Robertson et al., Phys. Rev. Lett. {\bf 67},
957 (1991).
\bibitem{11} H.S. Gurr, F. Reines and H.W. Sobel, Phys. Rev. Lett.
{\bf 28}, 1406 (1973).
\bibitem{12} A.P. Rekalo, Ukrain. Fiz. Zh. {\bf 18}, 213 (1973).
\bibitem{13} E. Majorana, Nuovo Cimento, {\bf 14}, 171 (1937).
\bibitem{14} P.B. Pal and L. Wolfenstein, Phys. Rev. {\bf D 25},
766 (1982).
\bibitem{15} B. Kayser, Phys. Rev. {\bf D 26}, 1662 (1982).
\bibitem{16} J.F. Nieves, Phys. Rev. {\bf D 26}, 3152 (1982).
\end{thebibliography}
\end{document}